\documentclass[12pt,preprint]{aastex} 
\slugcomment{}
\usepackage{graphics,graphicx}
\shorttitle{Redshifted Iron Lines}
\shortauthors{Turner, Kraemer, Reeves}

\begin{document}

\title{
Transient Relativistically-Shifted Lines as a 
Probe of Black Hole Systems }

\author{ T.\ J.\ Turner\altaffilmark{1,2}, S.\ B.\ Kraemer\altaffilmark{3}, 
 J.\ N.\ Reeves\altaffilmark{2} }

\altaffiltext{1}{Joint Center for Astrophysics, Physics Dept., University 
of Maryland Baltimore County, 1000 Hilltop Circle, Baltimore, MD 21250 }
\altaffiltext{2}{Laboratory for High Energy Astrophysics, Code 662, 
NASA/GSFC, Greenbelt, MD 20771}
\altaffiltext{3}{Catholic University of America, NASA/GSFC, Code 681, 
Greenbelt, MD 20771}

\begin{abstract} 

X-ray spectra of Seyfert galaxies 
 have revealed a new type of X-ray 
spectral feature, one which appears to offer important new insight into the  
black hole system. 
{\it XMM} revealed  
several narrow emission lines redward of  Fe K$\alpha$  in  
  NGC~3516. Since that discovery  the phenomenon has been observed in 
other Seyfert galaxies, e.g. NGC 7314 and ESO 198-G24. We 
present new evidence for a redshifted Fe line in {\it XMM} spectra 
of  Mrk~766. These data reveal the first evidence for a 
 significant shift  in the energy of such a line, 
occurring over a few tens of kiloseconds. This shift may be interpreted as 
deceleration of an ejected blob of gas 
traveling close to the escape velocity. 

\end{abstract}


\section{Introduction}

Active Galactic Nuclei (AGN) 
are believed to be powered   by  accretion of 
material onto a black hole.  UV photons from the disk are thought to 
be upscattered by relativistic electrons, providing the hard X-ray continuum.
These hard X-rays illuminate the disk surface, undergoing photoelectric 
absorption or Compton scattering. High abundance and fluorescence yield make 
  Fe K$\alpha$ the strongest line in 
 the X-ray band-pass, emitted via 
fluorescence or recombination processes  between  6.4 -- 7.0 keV,  
depending on the 
ionization-state of the gas. This line is commonly observed  
in AGN \citep{n97} with both narrow 
 and  broad components. The former has long been thought to be 
dominated by contributions from  cool material at or beyond 
the optical broad-line-region 
while the latter is thought to originate 
close to the black hole. Lines emitted very close to the black hole 
will show shifted, broadened and skewed profiles due to the combination 
of  Doppler and  relativistic effects (see \citealt{fab2000} and
references therein). Infalling material which is not part of the
 disk structure  will 
also suffer the effects of strong gravity, and may achieve 
velocities which are a significant fraction of the speed of light, causing
 significant displacement of  
emission lines from their rest energy. Many astrophysical sources 
show evidence for emission (SS~433, e.g. \citealt{mcs02}) and absorption from 
material traveling at relativistic velocities (e.g. Broad Absorption Line 
Quasars; \citealt{wey97}). 

Overlapping {\it Chandra}  and {\it XMM-Newton} 
 observations from 2001 November revealed  unexpected 
narrow 
emission lines at $\sim$ 5.6 and 6.2 keV in  the Seyfert galaxy 
NGC~3516 \citep{t02}. 
 Narrow emission  lines,  redshifted relative to Fe K$\alpha$ had never   
before been seen in the X-ray spectrum  of an AGN. 
The high-throughput of {\it XMM} and the excellent  energy-resolution 
afforded by the {\it Chandra} HETG made  possible the first unambiguous  
detection of  these  weak lines. 
With hindsight {\it ASCA} data show the appearance of narrow 
iron lines to be a frequently 
recurring phenomenon in NGC~3516 \citep{n99}   
 over a wide range in  source luminosity. 
\citet{t02} found  variability in the flux of the new lines   
on timescales of a few tens of ksec. 
The apparent variability of the 5.6 keV component was most interesting 
as it  could be attributed  either to the line flux varying or the 
line energy shifting from 5.4 to 5.6 keV. 
 A similar line at 5.86 keV  has now  been detected in NGC~7314 
\citep{y03}   and at 5.7 keV in 
ESO 198-G24 \citep{g03} -- the phenomenon  
may be a characteristic of AGN. 
Here we present data from an {\it XMM} observation of Mrk~766 which shows  
characteristically similar redshifted emission to that  observed in NGC~3516,  
offering new insight into the origin 
of these lines. 

\section{The case of Mrk 766}

{\it XMM} has three CCD detectors  (EPIC) 
operating in the $\sim 0.3 - 10$ keV band. The collecting area 
of the EPIC detector and telescope combination 
is approximately an order of magnitude higher than the 
CCDs flown on {\it ASCA}. The energy resolution is 
 $\Delta E \sim 120$ eV (FWHM) at 6.4 keV. This combination of capabilities 
has opened up  a new regime of Fe K$\alpha$ 
profile studies. High S/N spectra can be 
accumulated in much shorter integration times than ever before, allowing 
significant progress in studying variable spectral features and in 
detection of detailed structure to Fe line profiles. 
We analyzed the archival {\it XMM} data for Mrk 766. 
Two {\it XMM} observations of Mrk 766 are available in the archives, 
from 2000 May 20 and 2001 May 20-21. EPIC pn data were screened using 
SAS v5.3 software to 
select only events with patterns in the range 0 - 4. We also applied 
energy cuts to discard data below 0.2 keV and above 15.0 keV. 
The pn data yielded a useful exposure of 130 ksec for the 2001 observation and 
60 ksec for the 2000 data.  The 2001 
observation utilized the Small Window mode and there was no 
region for extraction of MOS2 background spectra, unfortunately 
this  background level 
becomes significant at and above the Fe K$\alpha$ emission line. The 
2001 MOS1 data were taken in timing mode, with the central chip 
off. In timing mode, spectra must be extracted based upon the histogram 
of counts versus RAWX coordinate. This allows the user to extract a spectrum 
for a given range of RAWX. While the histogram allows selection of 
source and background spectra, the source spectrum has to include 
some background dominated channels in the compressed spatial dimension. 
For weak sources such as Mrk~766 the resulting 
background-subtracted spectrum is 
somewhat inaccurate, especially at high energies. We extracted the 
timing-mode spectra for MOS1 and confirmed that those data were 
consistent with our PN results, but we do not present those data here.

In the 2000 observation  pileup hardened the MOS spectra but did not affect 
the pn data.  Thus the MOS data were not used from either observation. 
Source spectra were extracted from both epochs using circular extraction 
cells of $\sim 45''$ radii centered on Mrk 766.  Background spectra were 
taken from an offset position close to the source. 
Source spectra were grouped to 
have at least 20 counts per spectral bin for fitting using 
the $\chi^2$ statistic. Response matrices were generated for 
the source spectra using SASv5.3. 

\begin{figure}[!h] 
\includegraphics[scale=0.45,angle=0]{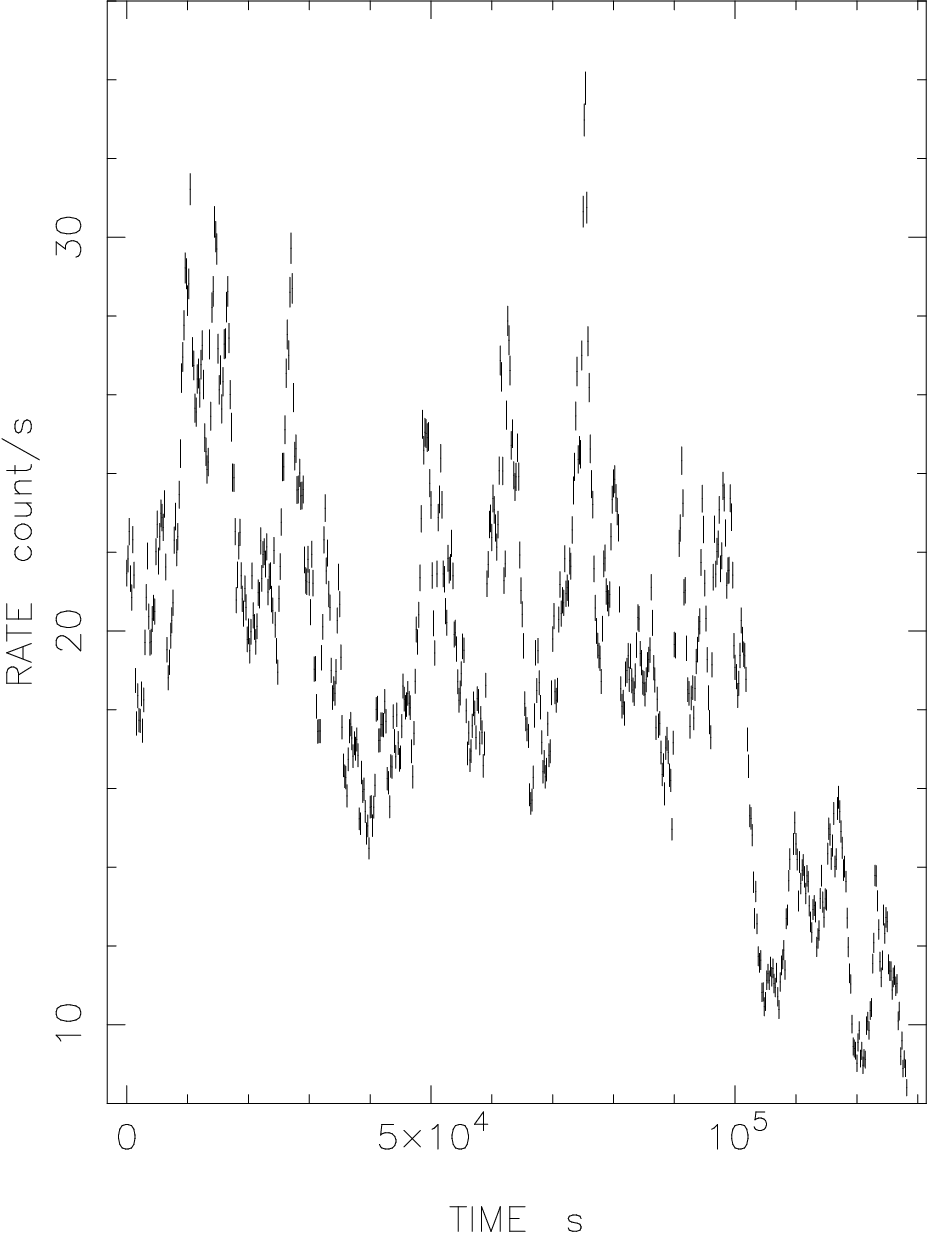}
\caption{\it Mrk 766: Light curve from the pn data taken during 
2001 May 20-21, accumulated from the source extraction cell 
in 200 second bins. The high-state corresponds to data from
the first 100,000 seconds, and the dip refers to the last 30 ks of
the observation where the source flux drops notably.
} 
\end{figure}

Figure 1 shows the light curve from 2001 May, as first presented by 
\citet{p03}. We confirm the general  findings of \citet{p03}  regarding 
the mean spectrum, i.e. 
that the spectrum is steep 
$\Gamma \sim 2.2$ and 
shows a broad component of Fe K$\alpha$ at rest-energy 
$\sim 6.7$ keV, with a narrower component at 6.4 keV. 
Our initial spectral fits   
parameterized the hard X-ray continuum 
in the 3-11.0 keV range using a powerlaw continuum form, 
but ignoring data in the 
5-7 keV band where the Fe K$\alpha$ line resides. We assumed 
a redshift z=0.012929 for Mrk~766 \citep{mcm}. 

\begin{figure}[!h] 
\includegraphics[scale=0.45,angle=0]{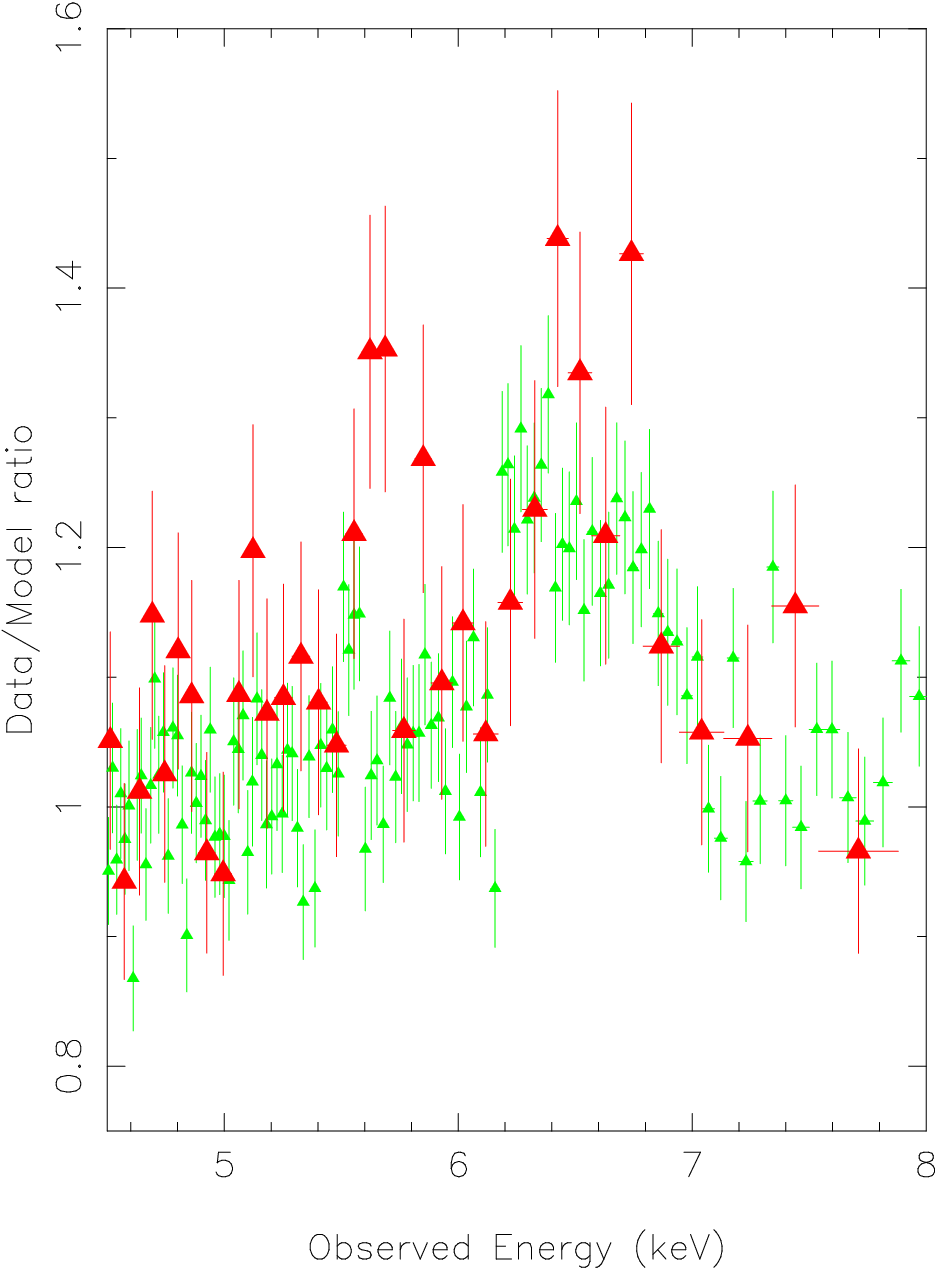}
\caption{\it Data/model of Fe K$\alpha$ emission in Mrk 766, where 
the model is the local powerlaw continuum, as detailed by \citet{p03}.  
The green line represents the high-state and the red line data from the dip. } 
\end{figure}

We also performed time-resolved 
spectroscopy,  accumulating spectra 
before (high-state)  and after (the dip) a drop in  source flux (Fig.~1). 
We adopted the general model of a powerlaw with 
broad and narrow components to model the complex 
 Fe K$\alpha$ line which is evident in the 6.4 -- 7.0 keV 
band. In addition, 
we find  a significant feature at 5.60 keV, not discussed by 
\citet{p03}. 

 Figure 2 shows the data/model line ratios 
for the two sections of data. An emission line is detected at 5.60 keV 
 in the high-state spectrum and at 5.75 keV in the dip-state. 
Both lines are detected at $ > 99\%$ confidence.  
The lines at 5.60 and 5.75 keV had photon fluxes 
 $n=0.50^{+0.26}_{-0.26} \times 10^{-5}$ photons cm$^{-2}$ s$^{-1}$ 
and 
$n=1.04^{+0.42}_{-0.42} \times 10^{-5}$ photons cm$^{-2}$ s$^{-1}$,  
respectively (errors represent 90\% confidence).  Subdivision of 
the high-state data into six time-resolved spectra of equal duration 
and the low-state into two halves 
revealed no other significant changes in line energy within each 
flux-state. 

We also  
compared the spectra of these two `states' 
with that from an {\it XMM} observation 1 year earlier. Figure 3  
shows the unfolded spectra from these three 
data sections. The apparent energy shift of the 5.60 keV feature 
is again evident by comparison of the top (green) and center 
(red triangles) lines. For comparison we 
include the 2000 May data (bottom line, black crosses), catching 
 the source at a flux state 
similar to that denoted the ``dip'' in 2001 May data. Those earlier data 
do not show a significant feature in the range 5.60 - 5.75  keV. 
The  2000 May spectrum yields an 
upper limit on the flux of a line at 5.75 keV 
$n < 0.2 \times 10^{-5}$ photons cm$^{-2}$ s$^{-1}$. Thus the source is 
inconsistent with the presence of the same line in both 2000 May 
and 2001 May spectra 
accumulated during the low flux-state. 

\begin{figure}[!h] 
\includegraphics[scale=0.45,angle=0]{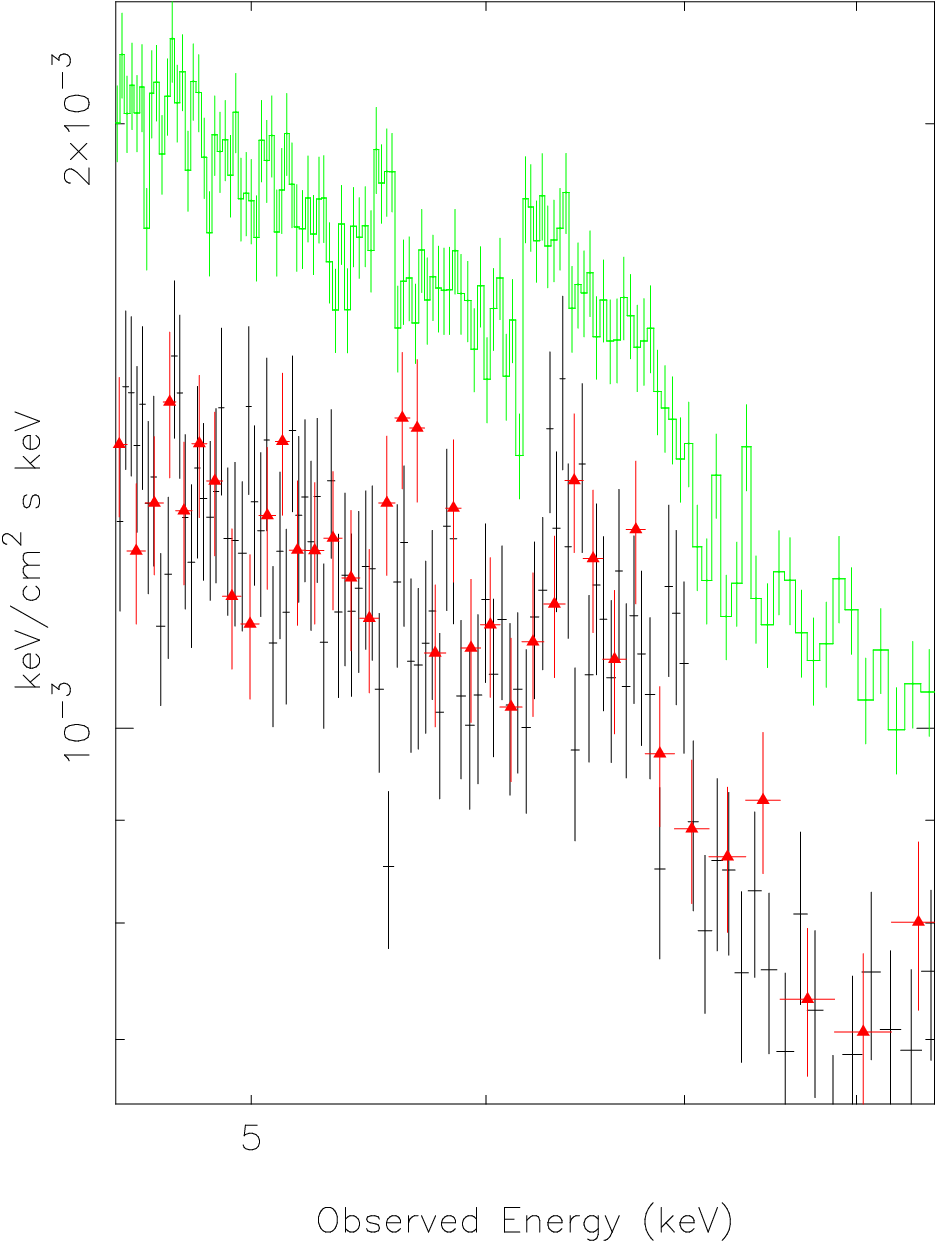}
\caption{\it Unfolded spectra from two XMM observations of Mrk 766. 
The top line (green) shows data from the high-state during the 
long observation of 2001 May 20-21. The red triangles   
show data from the dip-state of the same observation. 
A shift is evident in one line from 
approximately 5.60 keV to 5.75 keV within a few tens of ksec. 
The black crosses   
show data from the short observation of 2000 May 2000. 
} 
\end{figure}

To further investigate the narrow redshifted lines 
 we examined the significance of 
features in the Fe K$\alpha$ regime. We slid a Gaussian model template 
across data between 3.0 - 11.0 keV, testing for an improvement to the fit 
at every resolution element, compared to a powerlaw model for the 
underlying continuum. The background spectrum was subtracted before 
the test, so features in Figure 4 are attributable to Mrk 766. 

\begin{figure}[!h] 
\includegraphics[scale=0.45,angle=0]{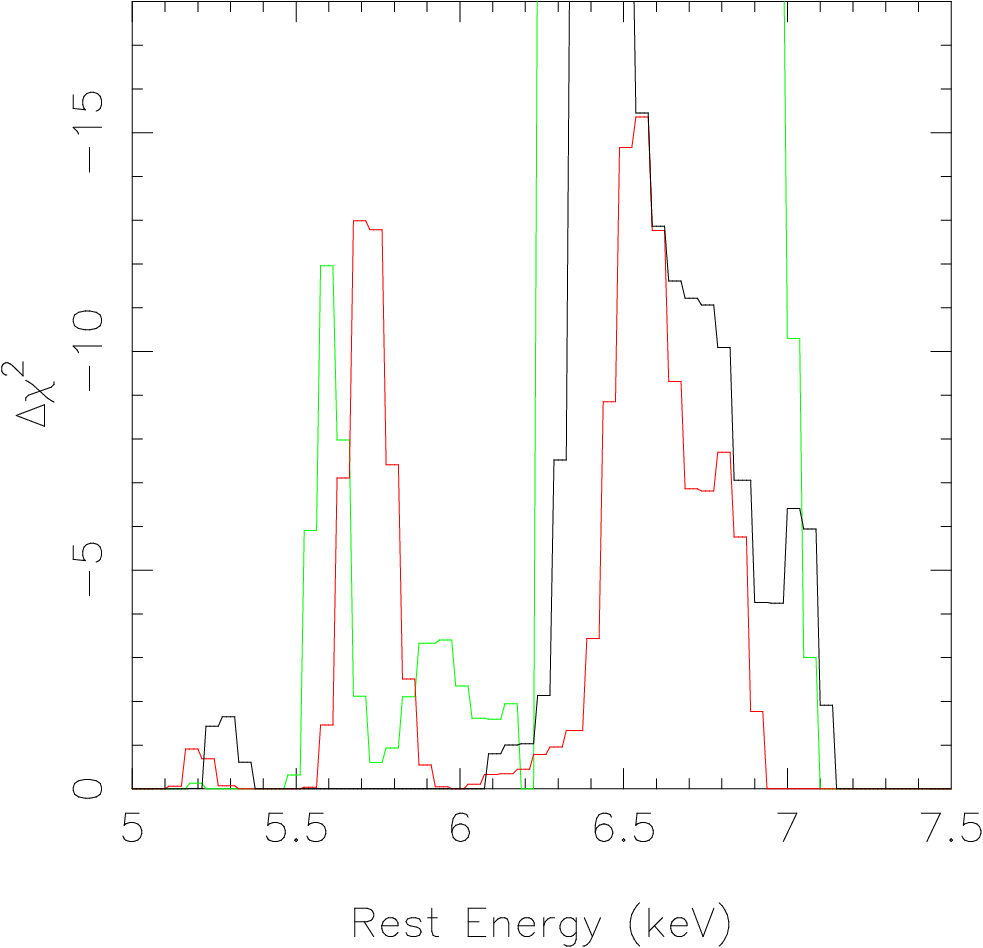}
\caption{\it Mrk~766 - 
Change in fit statistic  versus energy for a narrow Gaussian model line
which was stepped across the data. The comparison model was 
a simple absorbed powerlaw continuum, with  parameters allowed to 
float freely. The green line represents the high-state 
data from 2001 May, the red line is data from the dip and the black line is 
data from the earlier observation in 2000 May. The strongest Fe line component is also evident at 6.4 keV, the Y-axis is truncated to 
emphasize the weaker line.  } 
\end{figure}

In the resulting fit-statistic,  $\Delta \chi^2$ of 9.21 represents  
a feature which is significant at 99\% confidence. Fig.~4 
 illustrates  
the newly discovered  lines  and also traces out the stronger Fe lines 
reported by \citet{p03}.  
The green line represents the 2001 high-state, the red line 2001 dip data  
and the black line the 2000 May observation. 
Fig. 4  provides 
supporting evidence that the peak 
of the emission  
shifts from 5.60 to 5.75 keV in the rest-frame, and that there is no 
significant feature in the 2000 May spectrum. 
Figure 5 
 shows the contours for the energy and intensity of this line 
in the high and dip states. 
An apparent shift in line energy is significant at 95\% confidence (while 
the apparent change in line intensity turns out to be insignificant).  
Our analysis of a 90 ks {\it Chandra} HEG observation of Mrk~766 from another 
epoch did not show 
significant redshifted, narrow Fe lines.

\begin{figure}[!h] 
\includegraphics[scale=0.40,angle=0]{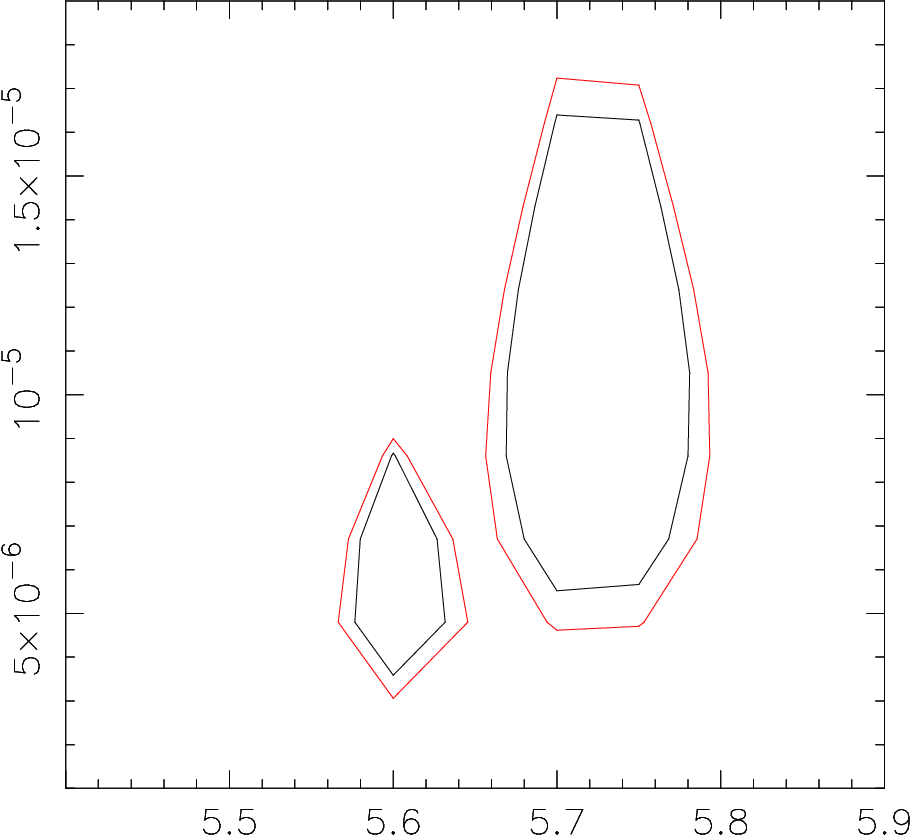}
\caption{\it Contours showing the  90\% (black) and 95\% (red) confidence 
levels for values of energy and intensity for the new line. 
The left-hand contours are from the high-state and the right-hand contours 
are from the dip, both from 2001 May.} 
\end{figure}

\section{Discussion}

Several cases of narrow and redshifted lines 
have now been found in Seyfert galaxies and  
 these features may be  a characteristic 
of  black hole systems, currently detectable 
only in the relatively bright Seyfert type AGN. 

Spectra of Mrk~766 show a narrow, redshifted Fe line which 
changes from 5.60 to 5.75 keV within a few tens of kiloseconds. 
These data provide the first ever observation  of a significant shift 
in energy for a line of this type. 
 We dismiss the possibility that the observed change in energy is due to 
an increase in ionization-state of the emitter. In that picture an 
increase in line energy should be correlated with an increase, 
rather than a decrease in central luminosity. 
 
Spallation, which  is the breakdown of Fe atoms into lower Z metals 
 by energetic protons,   
 has been  disfavored as an explanation of the lines 
in NGC~3516 \citep{t02} on the basis of line ratios and rapid 
line variability. The energy shift detected in the line from Mrk~766 
seems inconsistent with the spallation scenario, pushing it further from favor.

In the case of NGC 3516 \citet{t02} discuss an origin 
 from narrow radii which are sporadically illuminated 
on the surface of the accretion disk.   The 
``Thundercloud'' model may be applicable, where the 
accretion disk is illuminated in patches from magnetic reconnection events 
\citep{mf01}.  
If the redshifted lines observed in Mrk~766 are red horns  
of disklines then peak energies in the range 5.60-5.75 keV 
would occur from a radius $R \sim 30$ gravitational radii. 
For a disk inclined at 
30 degrees to our line-of-sight there should be an associated 
blue horn at $\sim 6.7$ keV, which would be $\sim 70\%$ 
stronger than the red horn due to relativistic effects 
(\citealt{schwarz}, \citealt{laor}). 
This is true for both the Schwarzschild and 
Kerr metrics, which produce similar lines at such radii. 
If the disk is inclined at 60 degrees then the blue horn would be 
expected at 7.25 keV with twice the strength of the red horn. The data 
are inconsistent with the latter line. 
Without an independent constraint on disk inclination 
it is difficult to make a conclusive comment on the consistency of the data 
with a blue horn and thus with the disk model. 
For the case of a disk inclined close to face-on 
the blue horn would be lost in the broad line component noted by 
\citet{p03}, or in the wings of the strong 
Fe K$\alpha$ component at 6.4 keV. 
A disk origin remains an interesting possibility. 
Shifts in line energy might be expected 
to be correlated with flux as new magnetic reconnections would likely  
light up different parts of the disk. 
Alternatively, the 
observation of a rapid  energy shift for the emission line in Mrk~766 
is suggestive of  
a decelerating blob of gas, and we now explore this possibility 
in detail.

\subsection{Outflow?}

That the line is redshifted could indicate either an origin in 
material expelled from the nucleus, in which case we are seeing 
material on the far side of the black hole, or material falling into
the nucleus, where we are viewing gas on the near side of the black hole. 
The observation that the 
line is getting less redshifted with time is indicative of 
deceleration of the emitter. Deceleration might be expected as  expelled 
material would be likely slowed by gravity. 
While there may exist deceleration mechanisms 
which can  slow infalling material (e.g. magnetic field pressure or 
drag forces) the outflow scenario 
seems a likely model 
 and so we restrict ourselves to a discussion of this possibility. 
NGC 3516 may show a similar deceleration, as there is some suggestion of 
a line which moves from 5.4 to 5.6 keV in a few tens of kiloseconds  
\citep{t02}. 
Understanding gas inflow/outflow is to get insight into 
the accretion/expulsion processes, two of the most fundamental questions  
regarding  black hole systems. 
Disk winds are a natural consequence of the 
accretion process, and so it is particularly compelling to look at these 
new observations in light of such models. 

One can use the observed shift in line energy to constrain the parameters of 
 gas ejected  ballistically from the disk surface. 
Consider a blob of gas propelled outward just above the plane of the disk. 
The propulsion could be driven in an number of ways, magnetically,  
or by the pressure gradient which must exist in the radial direction. 
The material is expelled and then decelerates under gravity. 

If the material is decelerating under gravity then 
$\Delta V/\Delta t = GM/R^2$, where $\Delta t$ is the time under which 
the deceleration was observed, and $R$ is the radius at which the material 
is situated at that time. Assuming the line  
originates from neutral iron then the implied initial velocity of the 
emitting blob is $38000$ km/s for the 
line at  5.60 keV in  Mrk 766.   
The observed change in velocity 
is  $\Delta V \sim 7000$ km/s in  $\Delta t \sim 65000 s$.  

Taking an estimate of the mass of the black hole to be 
 $10^7 M_{\odot}$ for Mrk 766 \citep{w02} 
 yields an estimate of the radius to be 
 $R \sim 3.5 \times 10^{14}$ cm. The escape velocity at this  
radius is  28000 km/s, 
 close to the (ejection) velocity observed.  This ties in 
nicely with the suggested outflow picture as blobs would have to achieve the 
escape velocity in order to be ejected and blobs not achieving this 
velocity would fall back to the disk. Application of the 
same arguments to the tentative energy shift in NGC 3516 
yields a velocity of 47000 km/s, with  $\Delta V \sim 9400$ km/s 
compared to an escape velocity of 41000 km/s at approximately the 
same radius (and central mass $\sim 2.3 \times 10^7 M_{\odot}$ ). 

Interestingly, \citet{kp03} 
note that the outflow velocity for the X-ray absorber in PG 1211$+143$ 
is close to the escape velocity for that inferred radius, suggesting 
these are related phenomena, where viewing angle determines whether one 
sees the ejected gas in emission or absorption.  We also note 
evidence for outflow of gas with velocity $\sim 50,000$ km/s in 
PDS~456, where \citet{reeves} estimated a launch radius of $\sim 40$ 
gravitational radii  
for the absorbing gas, if the outflow velocity is approximately equal to the 
escape velocity. 

The apparent link between the energy of the redshifted line, and the source 
flux-state during the 2001 observation 
is intriguing. However, the lack of a comparable feature in the earlier 
observation  (2000 May) indicates there is no long-term relation 
between the line and 
source luminosity. One possible explanation 
is that the emitting cloud responds to 
luminosity changes on short timescales, but that blobs do not exist on  
long timescales. This conjures a slightly more complex picture
than noted above. The blob may be driven outwards 
by radiation pressure and maintain a constant velocity so long as 
that pressure is balanced by gravity.  
 When the radiation field drops to half the 
previous value we obtain $\Delta V/\Delta t = GM/2R^2$ yielding 
 $R \sim 1.75 \times 10^{14}$ cm, and an escape velocity of 40,000 km/s. 
In this picture the blob is initially 
traveling outwards at a constant velocity 
approximately equal to the escape velocity. The source flux then 
drops by a factor 2 and the blob decelerates, dropping 
below the escape velocity. The evolution of the blob then depends on how the 
central luminosity continues to vary. If the source does not 
brighten again, the blob will fall back onto the black hole. 
This picture is appealing, but does not explain why the flux changes 
{\it within} the high-state did not show significant line shifts, 
leaving it unclear whether this picture is indeed more compelling 
than the simpler case of a blob driven magnetically or by pressure 
gradients. 
It is clear from the above estimates that 
observations with a long baseline which allow us to monitor these lines 
across several large changes in central luminosity could allow us 
to distinguish between radiative and other mechanisms driving the 
ejecta. 

While an ejection model might be expected to yield both blue and 
redshifted emission lines, assuming gas is symmetrically ejected 
towards and away from us, the blue lines will be harder to observe 
in available data. The blueshifted lines from a blob on the near-side 
of the black hole, traveling outward at the same velocity, would
be observed at $\sim 7.3$ keV. Interestingly the data 
show a couple of high points at 7.3 keV (Figs.~2, 3) however 
there are no significant lines present (Fig.~4).   
Upper limits on the flux of  a narrow line 
are $n< 9.0 \times 10^{-7}$  photons cm$^{-2}$ s$^{-1}$ 
($EW < 5$ eV) and $n< 3.5 \times 10^{-6}$  photons cm$^{-2}$ s$^{-1}$ 
($EW < 28$ eV) for the high and low states, respectively. 
  The emitting gas is likely to have a column 
$\sim 10^{24} {\rm cm^{-2}}$ \citep{fgf} 
 and hence will absorb much of the fluorescent line emitted on the 
illuminated side of the blobs on the near side of the hole, whereas we 
view the illuminated face of the receding blobs directly. 
This makes it difficult to rule out symmetric outflow.  
Interestingly, 
 lines have been observed blueward of 6.4 keV in sources such as 
NGC 7314 \citep{y03}. 
In that  case there was ambiguity as to whether the 
lines were redshifted 
H-like and He-like species of Fe  or blueshifted lines from 
neutral Fe. 

In Mrk 766 the line at $\sim 5.60$ keV has mean equivalent 
width $15^{+6}_{-5}$ eV. The strongest component 
of Fe K$\alpha$, at 6.4 keV, likely has 
contributions from the optical  broad line clouds  
with  $\Omega/2\pi \sim 
0.05-0.3$ in Mrk~766 \citep{nl93}. Adding  
contributions from the narrow-line region and torus yields 
 $\Omega/2\pi \sim 0.4$.  The 
ratio of  strengths of the  components at 6.4 and 5.6 keV  
 is $\sim 2.5$. 
Thus the redshifted line  likely has 
 covering fraction $\Omega/2\pi \sim 0.15$. 
Taking  our derived radii and column densities, a mass 
 $\sim 10^{-4} M_{\odot}$ is lost in the blob. 
The kinetic energy 
in the outflow is approximately $10^{48} {\rm ergs}$, a few 
percent of the gravitational potential 
energy available in the disk at these radii. 

\citet{p03} found a similar covering factor for the variable warm absorber 
in Mrk~766, and suggested that gas could contribute to the He-like 
Fe K$\alpha$ emission. We prefer a neutral origin for our emission line 
(if the line originated in He-like Fe then we would expect some H-like 
Fe and so would expect a pair of redshifted lines). However, in  
general the \cite{p03} picture of flare ejecta is consistent 
with the new result presented here. The variable absorption may 
be seen in the component of gas on the near side of the flow, and emission 
is observed from the component on the far side.

\citet{lev} discuss line photon propagation in  bulk outflows from AGN 
where Thomson scattering in the divergent flow redshifts photons to 
produce the broad red wing observed on the line profile of most AGN. 
Similar arguments could be applied to produce the narrow redshifted lines 
observed in NGC 3516, Mrk 766 and other AGN, although in the latter case 
the rapid shift 
in line energy may require a more complex explanation than the 
decelerating emitter, suggested above. 

\section{Conclusions}

Data from several  Seyfert galaxies  show 
narrow, redshifted emission  most likely 
explained as Fe K$\alpha$ lines, shifted by relativistic effects. 
Mrk~766 provides the first detection of a significant shift in the energy of 
such a line. 
The timescale of the energy shift is 
 a few tens of kiloseconds, indicating a possible origin in  
blobs of gas expelled from the nucleus and then
 gravitationally decelerated. 
The nature of the ejection mechanism is unknown at this time, but 
this may be a newly detectable signature of black hole systems. 

\section{Acknowledgments}

We thank Ian George and the anonymous referee for useful 
comments which improved the paper. 
T.J.\ Turner acknowledges support from NASA 
grant  NAG5-7538.

\end{document}